\documentclass[a4paper, reprint, rsi, aip, graphicx]{revtex4-1}
\usepackage{graphicx}
\newcommand{\mz}{\ensuremath{m_S=0} }
\newcommand{\mpm}{\ensuremath{m_S=\pm1} }

\begin{document}

\title{Camera-limits for wide-field magnetic resonance imaging of a nitrogen-vacancy spin sensor}

\author{Adam M. Wojciechowski}
\affiliation{Department of Physics, Technical University of Denmark, 2800 Kgs.~Lyngby, Denmark}
\author{M\"ursel Karadas}%
\affiliation{ 
	Department of Electrical Engineering, Technical University of Denmark, 2800 Kgs. Lyngby, Denmark
}%
\author{Alexander Huck}
\affiliation{Department of Physics, Technical University of Denmark, 2800 Kgs.~Lyngby, Denmark}

\author{Fedor Jelezko}
\affiliation{ 
	Institute for Quantum Optics, Ulm University, Albert-Einstein-Allee 11, 89081 Ulm, Germany
}%

\author{Jan Meijer}
\affiliation{ 
	Felix Bloch Institute for Solid State Physics, University of Leipzig, 04103 Leipzig, Germany
}%

\author{Ulrik L. Andersen}
\affiliation{Department of Physics, Technical University of Denmark, 2800 Kgs.~Lyngby, Denmark}

\begin{abstract}
Sensitive, real-time optical magnetometry with nitrogen-vacancy centers in diamond relies on accurate imaging of small ($\ll 10^{-2}$) fractional fluorescence changes across the diamond sample. We discuss the limitations on magnetic-field sensitivity resulting from the limited number of photoelectrons that a camera can record in a given time. Several types of camera sensors are analyzed and the smallest measurable magnetic-field change is estimated for each type. We show that most common sensors are of a limited use in such applications, while certain highly specific cameras allow to achieve nanotesla-level sensitivity in $1$~s of a combined exposure. Finally, we demonstrate the results obtained with a lock-in camera that pave the way for real-time, wide-field magnetometry at the nanotesla level and with micrometer resolution.

\end{abstract}

\maketitle

\section{Introduction}
Over recent years there is a surging interest in application of the negatively-charged nitrogen-vacancy (NV) color centers in diamond \cite{Jelezko2002, Doherty2013} to precision sensing of temperature \cite{Kucsko2013} and electro-magnetic fields\cite{Dolde2011,Rondin2014}, owing to a high-sensitivity per unit volume of such sensors. The smallest magnetic field sensors consist of a single NV center and provide nm-scale spatial resolution \cite{Maze2008}, while bulk diamonds typically allow for sensing at micro- and millimeter length scales. A thin layer of NV centers, engineered close to the surface of a diamond \cite{Rabeau2006, Pezzagna2010}, may be used for two-dimensional mapping of the magnetic field by projecting the NV fluorescence onto a camera sensor\cite{Pham2012}. Spatial resolution in 2D mapping depends on the imaging system and sensor parameters. Sub-micrometer resolution can be achieved with high magnification and high numerical-aperture (NA) microscopes equipped with a common CCD/CMOS \footnote{CCD - charged coupled device, CMOS - complementary metal-oxide} camera \cite{Pham2011, Chipaux2014}. Wide-field imaging techniques were recently used to measure fields generated by thin films \cite{Simpson2016} and magnetic beads \cite{Smits2016a}, and to reconstruct the current flow inside integrated circuits \cite{Nowodzinski2015} or graphene sheets \cite{Tetienne2016}. 

Until now, wide-field magnetic field imaging was restricted to steady fields, with multiple (typically hundreds or more) image frames processed in order to fit the optically detected magnetic resonance (ODMR) spectrum with a sufficient signal-to-noise ratio (SNR). Complete spectrum analysis provides full information on the magnetic field vector, at the cost of long measurement and post-processing times \cite{Chipaux2014, Glenn}. In this work we focus on real-time imaging of small magnetic-field changes that shift the ODMR resonance by no more than its linewidth. With the microwave (MW) frequency tuned to the side of one ODMR resonance, a projection of the magnetic field vector onto a certain spatial direction can be directly imaged, with the time resolution set by the camera frame rate. Since typical NV ensembles exhibit a fluorescence contrast of a few percent only, a sensor offering high SNR ($\gg$10$^2$) is required. Large SNR cameras have been recently developed for measuring small fluorescence variations of voltage sensitive dyes\cite{Homma2009, Park2011}. Our approach utilizes a sensor developed for optical coherence tomography \cite{Beer2004}, which allows for phase-sensitive detection of a fluorescence signal and greatly alleviates difficulties associated with low ODMR contrast.
	
In this article, we discuss the basic limitations for the NV diamond magnetometer sensitivity that are imposed by the camera serving as a fluorescence detector. We compare several sensor types and the sensitivity of a magnetometer using continuous-wave (cw) ODMR is estimated. The article is organized as follows: principles of magnetic field sensing using ODMR are described in Sec.\ref{sec:magnetometry}, in Sec.\ref{sec:sensitivity} the camera-imposed constraints are discussed, results of our experiments with a lock-in camera are presented in Sec.\ref{sec:results}, and the findings are summarized in Sec.\ref{sec:conclusions}. 


\section{NV magnetometry using ODMR}
\label{sec:magnetometry}

\begin{figure}[tb!]
	\centering
	\includegraphics[width=0.9\linewidth]{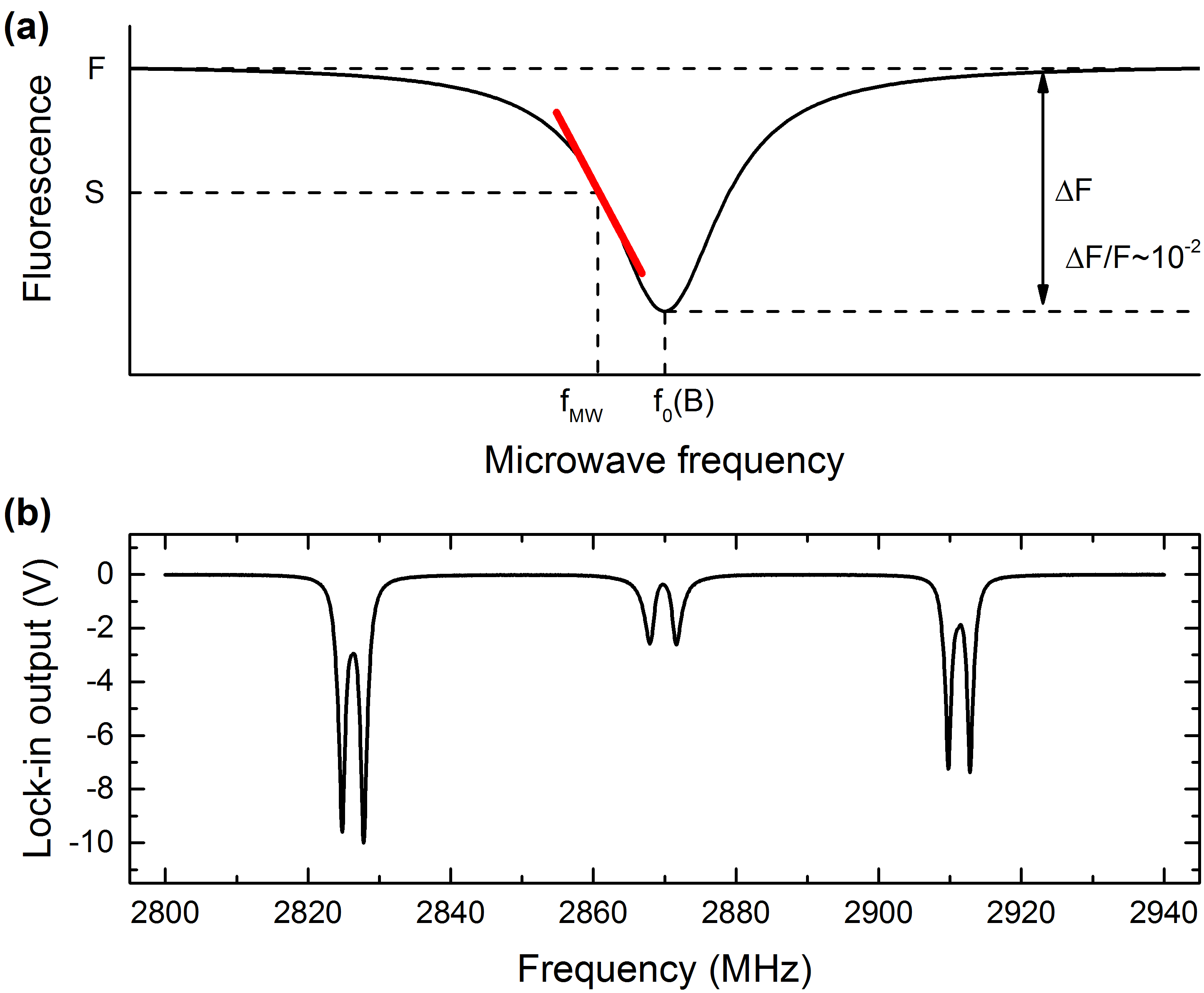}
	\caption{(a) Principle of magnetic field measurement using ODMR. For a fixed microwave frequency $f_{MW}$, the fluorescence signal $S$ is recorded. Small changes in the magnetic field proportionally shift the MW resonance position $f_0$. Red line highlights the resonance slope which is used to map the detected fluorescence variations with the frequency shift. (b) ODMR spectrum of $^{15}$NV recorded with a photodiode and amplitude modulated MWs in the bias magnetic field of $\sim$2~mT oriented in [110] direction. The peak depth corresponds to a contrast of $\sim$3\%.}
	\label{fig:odmr}
\end{figure}

The most common scheme for measuring dc and slowly-varying magnetic fields with NV centers relies on cw-ODMR. When the diamond sample is illuminated with green light, typically at a wavelength around 532 nm, a photoluminescence signal of the NV centers can be recorded in the $\sim$600-800 nm spectral range at room temperature. Simultaneously, NVs undergo a spin polarization into the \mz state within the $S=1$ ground state manifold, which provides higher fluorescence compared to the \mpm states. In the presence of a MW field resonant with the \mz $\longleftrightarrow$ \mpm transition, part of the \mz population is transferred back to the \mpm state, which is accompanied by the drop $\Delta F$ in the fluorescence level $F$, as schematically shown in Fig. \ref{fig:odmr}(a). Multiple MW resonances around the zero-field splitting of $\sim 2.8$~GHz can be observed in a typical ODMR spectrum, corresponding to the four possible orientations of the NVs in a diamond matrix and two (three) nuclear spin projections (hyperfine structure) of the $^{15}$N ($^{14}$N) atom constituting the NV center. With the magnetic field aligned to certain crystal directions, some spin transitions may become degenerate reducing the overall number of resonances in the spectrum. This can be seen in Fig.\ref{fig:odmr}(b), where magnetic field is aligned with the diamond [110] direction and hence only six resonances are visible. 
 
The exact resonance frequencies in the ODMR spectrum are determined by a projection of the magnetic field vector onto the NV axis. By recording a complete spectrum with ensembles of NVs pointing along all four possible directions in the diamond matrix, full information on a magnetic field vector can be retrieved. However, if only a single component of the vector is of interest, the diamond sample can be oriented in a way that maximizes the field projection along a particular NV axis. This results in a largest resonance frequency shift for a given field change, i.e. highest magnetic field sensitivity. Apart from sample orientation, the magnetic field sensitivity depends on the contrast $C = \Delta F/F$ and linewidth $\Gamma$ of the observed ODMR resonance. The contrast is typically of the order of 0.1-1\%, due to a finite difference in the fluorescence of \mz and \mpm states and the fact that only a fraction of NV ensemble is interacting with the MW field at a time. Slightly higher contrast ($\sim$10\%) is observed in diamond samples where NV centers were preferentially aligned along one crystal direction during its growth \cite{Edmonds2012,Lesik2014}. On the other hand, the resonance linewidth depends on the quality of diamond sample and reflects its spin-dephasing time $T_2^*$. The latter is strongly affected by the concentration of $^{13}$C and other impurities (including N and NV) in the diamond lattice. 

With the MW frequency tuned to the slope of an ODMR resonance [see Fig.\ref{fig:odmr}(a)] one may record fluorescence changes caused by the magnetic field variations. As the resonance frequency shifts, the observed fluorescence level changes accordingly to the resonance slope (red line). The sensitivity of such a measurement is fundamentally limited by the spin-projection noise:

\begin{equation}\label{eq:spinnoise}
	\delta B \simeq  \frac{\hbar}{g \mu_B} \frac{1}{C \sqrt{n_0 V \tau T_2^{*}}}, 
\end{equation}
where $g$ is the Land\'e factor, $\mu_B$ is the Bohr magneton, $n_0$ is the NV density, $V$ is the sample volume, and $\tau$ is the measurement time, $\tau >>T_2^*$. For simplicity we have omitted the numerical factor on the order of unity, resulting from the Lorentzian shape of the resonance, and we assumed that the magnetic field direction is aligned with the NV axis. The above equation can be understood as an uncertainty resulting from performing independent measurements on an ensemble of $n_0 \cdot V$ spins, repeated every $T_{2}^*$ time over the full measurement time of $\tau$. The presence of contrast $C$ in the denominator accounts for limited discernibility of the spin states by means of the fluorescence measurement. For realistic values of NV concentration being 1 ppm, contrast of $C=5$\% and a dephasing time $T_2^{*} = 1\,\mu$s, the resulting sensitivity limit is as low as $\delta B \sim 300\, \textrm{pT} \cdot \textrm{Hz}^{-1/2}$ for a sensing volume of 1 $\mu$m$^3$. An order of magnitude higher NV concentrations have been reported in \cite{Acosta2009} as well as longer dephasing times have been observed\cite{Maurer2012}. 

The spin-projection limit, given by Eq.~(\ref{eq:spinnoise}), holds when all fluorescence photons are detected. In the experiments, however, NV magnetometers suffer from a finite collection efficiency and transmission through the imaging optics, light trapping inside the diamond due to total internal reflection, and detector imperfections. Specifically, most camera sensors are subject to saturation and therefore allow only a limited number of photons to be recorded per unit time. It is thus sensible to formulate the \emph{practical} sensitivity in terms of the actual signal, $S$, and noise, $N$, ratio in the recorded image: 
\begin{equation}\label{eq:snr}
\delta B \simeq  \frac{\hbar}{g \mu_B} \frac{\Gamma}{C \cdot \mathrm{SNR}}, 
\end{equation}
where SNR~$\equiv S/N$. In general, the noise term consists of optical shot-noise $N_{opt}$ associated with the recorded fluorescence, electronic noise of the detector $N_{el}$, and other technical contributions. For an optimized (shot-noise limited) detector, $N_{opt}$ becomes a dominating noise term, $N\approx N_{opt}$, and the signal to noise ratio is simply given by SNR~$= \sqrt S$. The magnetic field sensitivity can then be improved by increasing the number of photons collected in the measurement.

\section{Sensitivity limits with camera detection}\label{sec:sensitivity}

\begin{figure}[tb!]
	\centering
	\includegraphics[width=0.9\linewidth]{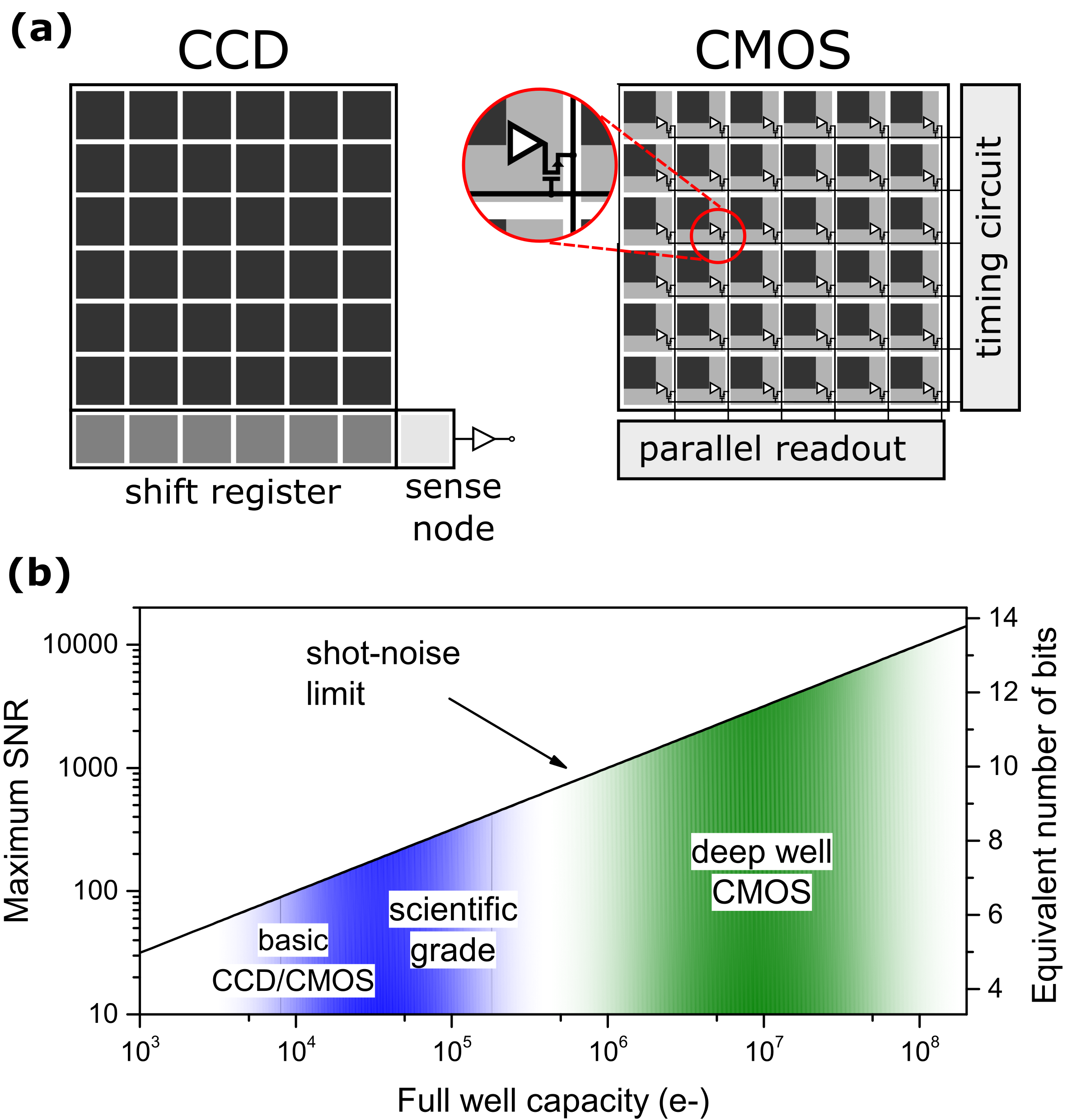}
	\caption{(a) Illustrative diagram of camera sensor operation. Photo-active area is shaded black. For a CCD matrix, the collected charge is transferred downwards line-by-line to the shift register and then consecutively converted to voltage and digitized. CMOS sensors use in-pixel transistors for charge conversion, which enables high-speed, random-access readout at the cost of a limited photo-active surface. (b) Maximum achievable SNR as a function of a pixel FWC. Left shaded area covers most common camera sensors. Right axis indicates the minimum ADC resolution needed to resolve pixel value without compromising SNR.}
	\label{fig:ccdcmossnr}
\end{figure}

The important ingredient for achieving high magnetic field sensitivity, as discussed above, is the ability of the imaging system to collect a large number of photons. In case of NV imaging this involves optimizing the delivery of excitation light, maximizing fluorescence collection efficiency and imaging optics throughput, and, finally, choosing a sensor capable of recording most of the incoming photons. The latter process is limited by the quantum efficiency (QE) of the sensor used, the size of the photo-active area within a pixel [fill factor (FF), see Fig.\ref{fig:ccdcmossnr}(a)] and, most importantly, by the finite amount of charge that can be stored in a pixel without signal degradation, referred to as the full well capacity (FWC). Below we discuss the parameters of imaging sensors that are important for high-sensitivity, wide-field magnetic imaging. In order to support the analysis with realistic parameter values we have analyzed several cameras with various sensor types which are listed in Table~\ref{tab:sensors}. For the sake of brevity, we neglect intensity fluctuations of the excitation light. 


\begin{table*}[tb!]
	\begin{tabular}{cllcccccc}
		\hline	
		camera&	manufacturer 		& sensor& pixel size 	& array size 		& QE 			& FPS with & FWC  		& DAC 	\\
		\#   & and model 							& type 	& ($\mu$m) & (pixels) & (\%) & full resolution & (e$^-$) & (bits) \\
		
		\hline
		1 & Thorlabs DCC3240N 			& CMOS 	& 5.3$\times$5.3& 1280$\times$1024 	& $\sim60$		& 60 	& ~~$8 \times 10^3$ 	& 10 	\\
		2 & IDS UI-3140CP 				& CMOS 	& 4.8$\times$4.8& 1280$\times$1024 	& 30-55			& 224	& ~~$1 \times 10^4$	& 10 	\\
		3 & Andor iXon Ultra 897			& EM-CCD& 16$\times$16	& 512$\times$512	& $>90$ 		& 56	& $1.8 \times 10^5$	& 16 	\\
		4 & RedShirtImaging NeuroCMOS-DW128f & CMOS& 128$\times$128& 128$\times$128 	& $\sim50$ 		& 10 000	& ~~$1 \times 10^8$ & 14 	\\
		5 & Heliotis heliCam C3			&lock-in& 39.6$\times$39.6	& 280$\times$292	& 60-80		& 3 800\footnote{internally up to 1 000 000 FPS (modulation frequency $\leq $ 250 kHz)}	& $3.5 \times 10^5$ & 2$\times$10\footnote{separate in-phase and quadrature output images} \\
		\hline
	\end{tabular}
	
	\caption{Camera sensor parameters used for determination of achievable SNR and magnetic field sensitivity.}
	\label{tab:sensors}
\end{table*}

\paragraph{Full Well Capacity.}
When the camera is overexposed, i.e., the number of photoelectrons exceeds the FWC, then the recorded image becomes distorted. The exact signal degradation depends on the sensor technology and particular design. However, the main sources of distortion are the saturation of intensity recorded by a pixel and the leakage (bleeding) of the photoelectrons to the neighboring pixels, a process called blooming and affecting mostly CCD sensors. During a single exposure of the camera, the maximum number of photoelectrons (charges) captured by a pixel is $S_{max}=$~FWC, in order for the image to be recorded accurately. This number is independent on the fluorescence collection efficiency $\eta$, exposure time $\tau$, pixel size, quantum efficiency (QE) or filling factor (FF) of the sensor itself. The optical shot-noise, $N_{opt}$, associated with such a signal level also reaches the maximum value $N_{opt, max} = \sqrt{\mathrm{FWC}}$, and dominates other noise types. The readout noise is typically on the order of 1-10 e$^-$ and is relevant only for low light levels, while the dark-noise affects long-exposure images. Therefore, the signal-to-noise ratio per pixel and a single camera exposure is equal to SNR$_{px}^{exp} \approx \sqrt{S}$ and has a maximum value of: 
\begin{equation}\label{eq:snr_px_exp}
	\mathrm{SNR}_{px, max}^{exp} \approx \sqrt{\mathrm{FWC}},
\end{equation}
when the camera is fully exposed. Since each pixel can collect only up to $S=$~FWC photoelectrons, Eq.~(\ref{eq:snr_px_exp}) sets an upper limit on the signal-to-noise achievable for a given sensor design, independently from the optical imaging setup efficiency. Figure \ref{fig:ccdcmossnr}(b) illustrates the maximum achievable, shot-noise limited SNR for a given capacity of a sensor pixel. Typical camera sensors have a FWC in the 10$^4$-10$^5$ range, with SNR $\lesssim300$. Sensors with a largest FWC are preferable for sensitive ODMR-based magnetometry, provided the number of fluorescence photons arriving during a single frame exposure time is sufficiently high.

\paragraph{Camera framerate}

For a given well capacity of a pixel, the number of photoelectrons collected per unit time can be maximized by running the camera at high frame rates (under sufficiently strong illumination). The maximum number of frames a camera is able to capture per second (FPS) is a parameter directly specified by the manufacturer and the maximum number of photoelectrons recorded in 1 s, $S_{px}^{1s}$, is given simply by $S_{px}^{1s} = \mathrm{FPS}\cdot \mathrm{FWC}$. Although the camera has a certain dead (processing) time it does not limit the sensitivity as long as there's enough light available for exposure. The bandwidth-normalized SNR is thus given by:
\begin{equation}\label{eq:snr_px_1s}
	\mathrm{SNR}_{px, max}^{1s} = \sqrt{\mathrm{FPS}\cdot \mathrm{FWC}}.
\end{equation}
The above equation shows that the best camera is the one capable of collecting most photoelectrons per frame and operating at the highest frame rate. For the deep-well CMOS sensor \#4 (see Table~\ref{tab:sensors}), in order to achieve the maximum sensitivity a pixel has to collect as many as $10^{12}$ photoelectrons per second, which amounts to $\sim$1~$\mu$W of optical power delivered to a pixel and $\sim$10~mW in total for uniform sensor illumination. Although these are high fluorescence levels, even larger values from a quasi-2D NV layer have been reported in Ref.\cite{Barry2016}. 


\paragraph{Sensor resolution}

Most common laboratory cameras offer resolution in the megapixel range, and use sensors of a (1/3)" to 1" size, which have a side length of the photosensitive area $\sim $5-13~mm. That implies a single pixel length is on the order of $\sim$1-10~$\mu$m. The larger the pixel size the higher is its FWC and the more photons can be collected before saturation. As a consequence, sensors with largest potential wells usually have a reduced spatial resolution, as can be seen in Table~\ref{tab:sensors}. However, high-resolution cameras offer the additional possibility to combine data from several small pixels to form a single 'macro' pixel. This can be done through post-processing of image data, with an effective FWC of a macro pixel being the sum of FWCs of constituting pixels, at a cost of larger amount of data required to be processed. Many sensors offer direct binned readout, e.g., \#1 and \#2. The on-sensor binning, however, doesn't improve the effective FWC. 

Another important factor for accurate imaging is the digital resolution of the analog-to-digital converters (ADC) in the camera. Entry-level cameras typically output 8-10 bit images (for color cameras separately for red, green and blue components), while the scientific grade cameras most often output 12-16 bit images. The ADC resolution is usually matching the ratio of the FWC to the electronic readout noise. This value defines the camera dynamic range, i.e., ratio of the largest to the smallest detectable signal. However, in order to effectively use the large FWC for precision ODMR detection, the number of bits needed to capture images with the maximum SNR is given by $\log_{2}(\mathrm{SNR}_{px}^{max}) = 0.5\log_2(\mathrm{FWC})$ and summarized in Fig. \ref{fig:ccdcmossnr}(b). All cameras listed in Table~\ref{tab:sensors} provide enough ADC resolution for that purpose. Additional bits are useful for exploiting the full dynamic range of the sensor, but come at a cost of a larger bandwidth required for continuous image data transfer or the necessity of an internal camera memory for a burst-mode video capturing.

\paragraph{Quantum efficiency and fill factor}

The number of photoelectrons collected by the camera forms a fraction of all incoming photons due to the finite quantum (conversion) efficiency and a limited photo-active area within each pixel. Due do their design, CCD cameras offer almost unity FF [Fig. \ref{fig:ccdcmossnr}(a)] and highest QE, exceeding $>90$\% for back-illuminated sensors. On the other hand, most CMOS cameras have a limited FF, due to the presence of several transistors in each pixel [Fig. \ref{fig:ccdcmossnr}(a)]. This is often mitigated by the inclusion of micro-lens arrays in CMOS cameras, which focus light onto the photo-active regions and lift the effective FF close to unity. 
For both technologies, QE$\cdot$FF exceeds $0.5$ for the best cameras and, therefore, these parameters do not influence the sensitivity by much. 

\paragraph{In-pixel lock-in detection}

With the advent of modern CMOS fabrication processes, a new type of cameras became commercially available, so called time-of-flight (ToF) cameras, with their primary use in range-finding \footnote{one example is the Microsoft Kinect camera}. In the simplest ToF sensor, the charge accumulated on a pixel photodiode is transferred by accurately-timed gating electronics to one out of two holding wells (capacitors). Arrival time of a light pulse may be then determined by comparing the number of photons (collected charges) that arrived before and after the trigger that switches the active capacitor.
The concept of the ToF camera has been extended onto a phase-sensitive  demodulation. The charge from a pixel photodiode is transferred sequentially between four wells, with the cycle period matching the applied modulation frequency \cite{Spirig1997, Oggier2004}. In such a case, the voltage measured across the first (second) and third (fourth) well represents an in-phase (quadrature) signal. Low-pass filtering is performed by accumulating the charges over many modulation periods before reading out the pixel value.  This enables per-pixel lock-in demodulation of the optical signal, which is the main operation principle behind the \#5 lock-in camera. This camera sensor can additionally, after each modulation cycle, perform the dc signal (common-mode charge) subtraction and, therefore, can more effectively use the FWC of a pixel, as will be discussed below. 

With full amplitude modulation of the MW field, the ODMR signal consists of the steady fluorescence level $S$ and a modulated part of $C S$ amplitude. When using the dc voltage subtraction feature of the sensor, the effective FWC is reduced by half\footnote{Heliotis AG, Switzerland, private communication (2016)}. The maximum signal that can be recorded is then given by the relation: $C S = \mathrm{FWC}/2$ or, equivalently, $S = \mathrm{FWC}/2 C$. The latter form, allows us to determine optical shot noise level to be $N=\sqrt{\mathrm{FWC}/2 C}$, which includes the noise from photons corresponding to the subtracted dc part of the signal. The magnetic field sensitivity for such a camera is, therefore, given by:
\begin{equation}\label{eq:snr_lockin}
	\delta B \simeq  \frac{\hbar}{g \mu_B} \frac{\Gamma}{C S/(\sqrt{2}N)} = \frac{\hbar}{g \mu_B} \frac{\Gamma}{\sqrt{\mathrm{FWC}\cdot C/4}}, 
\end{equation}
where the noise has been multiplied by factor of $\sqrt{2}$ that comes from the output signal being voltage measured across 2 capacitors with similar noise. Equation~(\ref{eq:snr_lockin}) indicates that the lock-in camera offers a sensitivity improvement by a factor of $1/\sqrt{4 C}$ over a non lock-in sensor with an identical pixel FWC, which is of the highest importance for low-contrast diamond samples. Additionally, the phase sensitive detection reduces the technical noise, e.g., originating from green light intensity fluctuations, by means of a common-mode rejection.


\begin{table}[t!]
	\begin{ruledtabular}
	\begin{tabular}{ccccc}

		camera & SNR$_{px, max}^{exp}$  &  $\delta B_{px, max}^{exp}$ & $\delta B_{px, max}^{1s}$ &  optical   \\
		  \#   &   &  (nT)      &  (nT)      &  power \\ \hline
		  1    &     89      &      7979      &     1030      &   0.3 $\mu$W \\
		  2    &     100     &      7137      &      477      &   2.2 $\mu$W \\
		  3    &     424     &      1682      &      225      &   0.9 $\mu$W \\
		  4    &   10 000    &       71       &      0.7      &  10.0 mW	\\
		  5    &     66\footnote{due to a different operation principle, a non lock-in camera requires SNR$_{px, max}^{exp}=1323$ to achieve the same field sensitivity.}    &      539       &      8.8      &  1.0 mW \\ 
	\end{tabular}
	\end{ruledtabular}
	\caption{Maximum SNR achievable with a single pixel and the corresponding magnetic field sensitivity estimated for a single frame and for one second at the maximum camera frame rate, respectively. Last column indicates the optical fluorescence power that is required to achieve $\delta B_{px, max}^{1s}$ sensitivity over the full sensor area.}
	\label{tab:sensitivity}
\end{table}

\paragraph{Magnetic field sensitivity estimation}

Using the parameters of cameras shown in Table~\ref{tab:sensors}, Eq.~(\ref{eq:snr}) and SNR formulas discussed above we estimate the values of a minimum detectable magnetic field that can be resolved by each camera. For simplicity, we determine the sensitivity of a single pixel (neglecting possible binning, and averaging over several pixels) for a single frame and for 1 s combined exposure (using maximum frame rate supported with full camera resolution). We assume here an ODMR contrast $C=5\%$, the resonance width $\Gamma/2\pi = 1$~MHz and that the magnetic field is aligned with the NV axis. Results of the sensitivity estimation are shown in Table~\ref{tab:sensitivity} together with the fluorescence light power required (assuming $\lambda = 650$~nm) in order to sufficiently illuminate the full sensor. 

\section{Experimental results}
\label{sec:results}

Our diamond sensor is a 2$\times$2$\times$0.5~mm$^3$ electronic-grade diamond substrate (Element 6), on top of which a $^{12}$C isotopically purified and $^{15}$N nitrogen-rich layer ([$^{12}$C]$>$99.99\%, [$^{15}$N]$\sim$10~ppm, 1~$\mu$m thickness)  was grown. Vacancies were introduced by 1.8~MeV helium ion implantation ($\sim$10$^{15}$~cm$^{-2}$ dose) followed by 2~h annealing in vacuum at 900$^\circ$C. The resulting NV concentration is estimated to be on the order of 1 ppm. 

Light is delivered to and collected from the diamond through the same microscope objective (Mitutoyo M Plan Apo NIR HR 100x, NA=0.7) in an inverted microscope arrangement. The top diamond surface is coated with a 300-nm-thick aluminum layer to reflect the excitation and fluorescence light. Te diamond bottom side is anti-reflection coated with silica, further aiding the light collection and reducing interference within the diamond. Typically, around 5~$\mu$W of red fluorescence is detected for 150~mW of excitation light (Verdi G8, 532 nm). The illuminated area is imaged on a camera sensor with 45$\times$ magnification, leading to a pixel size of $\sim$0.9 $\mu$m at the diamond plane. The actual resolution is lower ($\sim$2~$\mu$m), as the objective is not compensated for imaging through the diamond. MWs were generated with a signal generator (Stanford Research Systems SG394) and TTL modulated with a switch (Mini-Circuits ZASWA-2-50DR+). The lock-in camera (Heliotis heliCam C3) provided a square-wave modulation signal with a frequency set to 3.7~kHz. After passing the switch, MWs are amplified by a high-power amplifier (Mini-Circuits, ZHL-16W-43+) and delivered to the printed circuit-board antenna \cite{El-Ella2017} placed under the 3D-printed diamond holder with a distance of 1~mm between the copper trace and the NV sensing layer. A pair of permanent magnets created a uniform magnetic field of around 2 mT parallel to the diamond surface and aligned with the [110] direction.    

\begin{figure}[tb!]
	\centering
	\includegraphics[width=\linewidth]{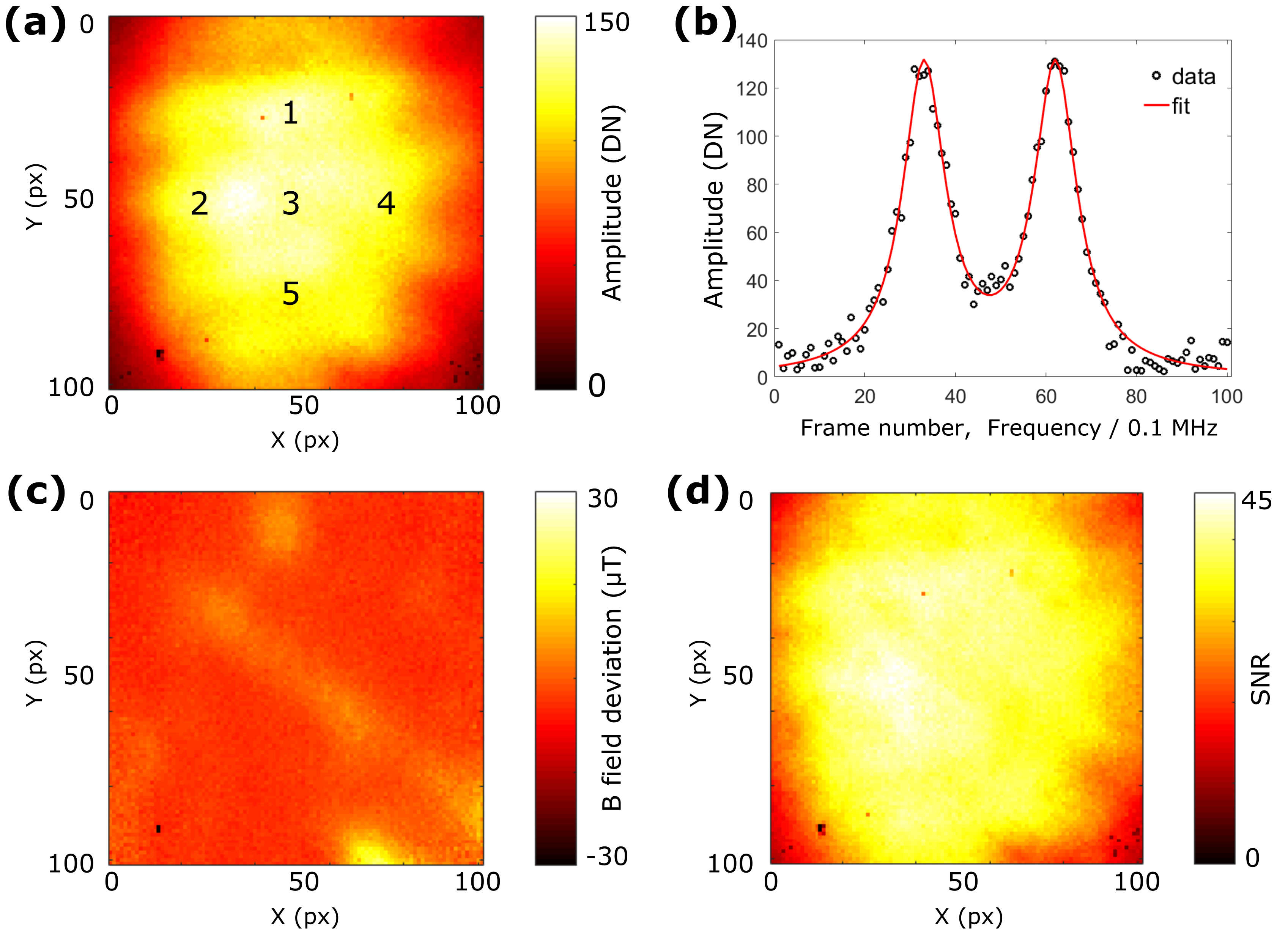}
	\caption{Magnetic field imaging using the lock-in camera. (a) 100$\times$100 pixel fluorescence intensity profile. DN refers to data number (ADC units) (b) Central-pixel ODMR spectrum taken from 100 frames captured during 10-MHz-wide MW sweep in a uniform magnetic field. (c) Relative position of the resonance frequency determined from the fits and plotted in magnetic field units. (d) SNR profile calculated from the fit parameters and noise amplitude measured without MWs.}
	\label{fig:heliotis}
\end{figure}

Figure \ref{fig:heliotis} shows the results of fluorescence and ODMR measurements using camera \#5. All images were captured in full resolution, however, only the 100$\times$100 pixels region with the illuminated spot is shown. The fluorescence profile recorded with the camera operating in the intensity mode, i.e., when no phase-sensitive detection is performed, is shown in Fig.\ref{fig:heliotis}(a). The deviation from Gaussian-shape results from the imperfection of the excitation beam-profile as well as non-uniform nitrogen incorporation during the CVD growth of the sensing layer and residual interference within the diamond. Five points in a cross-like arrangement around the center were selected for the sensitivity analysis, as indicated in Fig.\ref{fig:heliotis}(a). 

For the phase-sensitive demodulation, the camera was set to accumulate the signal over 62 modulation periods with background subtraction enabled, resulting in the frame rate FPS$=59.7$. ODMR spectra were recorded by sweeping the MW frequency over a 10 MHz range synchronously with the acquisition of 2$\times$100 frames, where the factor of 2 comes from simultaneous acquisition of in-phase (I) and quadrature (Q) images. The two quadratures were subsequently combined for each (x,y) pixel via $S(x,y) = \sqrt{I(x,y)^2+Q(x,y)^2}$. Figure \ref{fig:heliotis}(b) shows the ODMR resonance data recorded at the central point, S(50,50). Two hyperfine resonances separated by 3 MHz are visible, corresponding to the left pair of resonances shown in Fig.\ref{fig:heliotis}(b). The phase-sensitive demodulation of the signal results in zero background. In our case, however, the signal is calculated from both quadratures and the presence of noise translates into a small background shift. For each pixel a double-Lorentzian curve was fitted, assuming the same amplitude and width for both peaks. The result of a fit for the central pixel is shown in Fig.\ref{fig:heliotis}(b). The resonance amplitudes at points 1-5, extracted from the fit parameters, were equal to 142.4, 126.4, 133.0, 122.8, and 116.1, respectively. Spectral position of the lower-frequency peak, determined from the fit, was used to calculate the magnetic field variation across the full image. The resulting field map is shown in Fig.\ref{fig:heliotis}(c) and is in agreement with a diamond being placed in a uniform magnetic field. Only small ($<$5~$\mu$T) variations are visible in the image except for a larger deviation ($\sim$15~$\mu$T) spot at the bottom, where, however, the fluorescence intensity is low. 

To quantify the SNR in the recorded data, an additional set of images was taken in the absence of MWs. These were similarly converted to amplitude images, and the standard deviation across 200 frames was calculated independently for each pixel. Fitted ODMR peak amplitudes and noise data were then used to create the SNR map shown in Fig.\ref{fig:heliotis}(d). In the central region, SNR exceeds 40 which translates into 1~$\mu$T magnetic sensitivity in a single-frame and a bandwidth-normalized sensitivity of 142~nT~Hz$^{-1/2}$. The latter is limited by the frame rate at which we operate the camera and, in-principle, could be improved by up to 8 times if more fluorescence light was available.  

\begin{figure}[tb!]
	\centering
	\includegraphics[width=\linewidth]{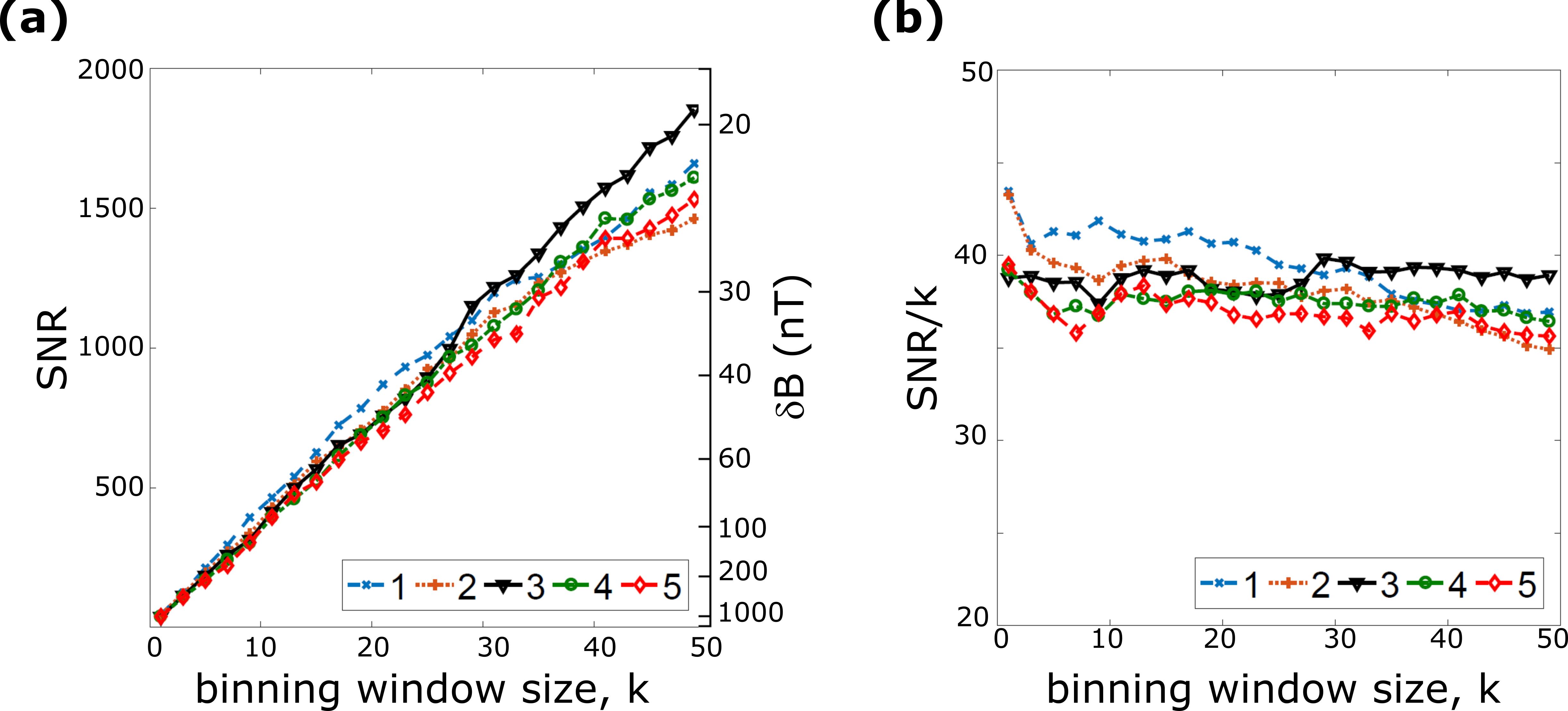}
	\caption{(a) SNR calculated for macro-pixels consisting of k$\times$k pixels around positions indicated by labels 1-5 in Fig.\ref{fig:heliotis}. (b) SNR data normalized to the window size. For small bins, the SNR is determined by the intensity profile.}
	\label{fig:binned}
\end{figure}

When the spatial resolution of a camera can be sacrificed, additional gain in the sensitivity can be achieved by means of a binning of the pixel data. The amplitudes extracted from the fits described above where summed over a $k\times k$ pixel region around the points labeled in Fig.\ref{fig:heliotis}, and the corresponding noise level was calculated from the data in the absence of MWs. The resulting SNR is shown in Fig.\ref{fig:binned}(a). As can be expected for white uncorrelated noise, the SNR scales linearly with the \emph{macro} pixel length (square root of the area). For the bin size of 49 pixels a sensitivity of $\sim$20 nT is achieved, which corresponds to the bandwidth-normalized value of $\sim$2.6~nT~Hz$^{-1/2}$. An independent measurement using a photodiode and a standard lock-in amplifier yields a similar sensitivity of $\sim$1.5~nT~Hz$^{-1/2}$ of our setup, albeit all available light was collected by the diode. The SNR normalized to the bin size is shown in Fig. \ref{fig:binned}(b). The initial values (for low $k$) depend heavily on the pixel choice, as the recorded intensity profile is not uniform. In one region an SNR~$>40$ is maintained over a 20$\times$20 pixel area. As the binning window size is increased all values converge around 35. For $k>25$, the binning areas partially overlap resulting in a further loss of spatial resolution.

\section{Summary and conclusions}
\label{sec:conclusions}

Several conclusions can be drawn from our analysis and the data shown in Table~\ref{tab:sensitivity}. Firstly, standard CCD/CMOS sensors, with FWC $\sim 10^4$-$10^5$, allow for single-exposure sensitivity on the order of a large fraction of the resonance linewidth, as the signal to noise multiplied by the ODMR contrast is on the order of unity. Higher sensitivity can be achieved by temporal averaging, especially with fast frame rate cameras, and/or by binning the captured images, sacrificing the spatial resolution.  
On the other hand, deep well sensors allow for much higher sensitivity by allowing measurements of small relative fluoresce changes $\Delta F/F \sim 10^{-4}$, when sufficient amount of light is available. The required  fluorescence levels were recently reported in \cite{Barry2016} for a dense ([NV]/[C]$\sim$1~ppm), 200x2000x13 $\mu$m$^3$ size NV layer with a similar resonance linewidth as in our sample.

Lock-in sensors form an interesting choice of the imaging device for multiple reasons: the dc subtraction mechanism allows for efficient usage of the pixel well capacity; the pixel value after background subtraction may be directly proportional to the field change; and the technical noise due to the laser instability is easily suppressed. The latter becomes increasingly important when smaller fluorescence changes are to be measured at higher frame rates. 

We have experimentally demonstrated the single-pixel and single-frame sensitivity of around 1~$\mu$T while maintaining acquisition speed of 60 frames per second. There are several ways in which the sensitivity of our setup can be further improved. Firstly, by adjusting the demodulation phase such that only in-phase or quadrature image would be needed. This can be accomplished by  using an external demodulation reference and would result in a $\sqrt{2}$ sensitivity improvement. Secondly, the camera frame rate can be increased up to 64 times. This requires an improved fluorescence collection and would result in a further 8-fold improvement of the bandwidth-normalized sensitivity. Finally, multiple successive frames or nearby pixel values can be averaged to further increase the sensitivity, at a cost of reduced spatial or temporal resolution. 


In conclusion, we have discussed the main camera-imposed limitations for the precision, wide-field NV-fluorescence detection. We have also shown that the nanotesla-sensitivity magnetometry at video frame rates is possible. Such imaging systems could be useful for sensitive monitoring of small field changes in real-time and with high spatial resolution, opening new possibilities for   studying dynamical systems.  

\begin{acknowledgments}

We thank Christian Osterkamp for the growth of the nitrogen-doped layer, Steffen Jankuhn for performing ion implantation and Kristian Hagsted Rasmussen for coating the diamond. We acknowledge stimulating discussions with Haitham El-Ella, Sepehr Ahmadi and Axel Thielscher. This work was supported by the Innovation Fund Denmark (EXMAD project) and Novo Nordisk Foundation.

\end{acknowledgments}

\bibliography{../library}

\end{document}